\documentstyle[twocolumn,aps,epsf]{revtex}

\begin{document}
\draft
\title{Circuit model for spin-bottleneck resistance in
magnetic-tunnel-junction devices}
\author{T. Jungwirth$^{1,2}$ and A.H. MacDonald$^{1}$}
\address{$^{1}$ Department of Physics, Indiana University, Bloomington, IN 47405}
\address{$^{2}$ Institute of Physics ASCR, Cukrovarnick\'a 10, 162 00 Praha 6,
Czech Republic}
\date{\today}
\maketitle

{\tightenlines
\begin{abstract}
Spin-bottlenecks are created in magnetic-tunnel-junction devices 
by spatial inhomogeneity in the relative resistances for up and down 
spins.  We propose a 
simple electrical circuit model for these devices 
which incorporates spin-bottleneck effects and 
can be used to calculate their overall resistance and magnetoresistance.
The model permits a simple understanding of the 
dependence of device magnetoresistance on 
spin diffusion lengths, tunneling magnetoresistance, and 
majority and minority spin resistivities in the 
ferromagnetic electrodes. The circuit model
is in a good quantitative agreement with detailed 
transport calculations.
\end{abstract}
}


\narrowtext

Magnetic tunnel junction (MTJ) devices consist of two 
ferromagnetic electrodes separated by an insulating 
tunnel junction.  The resistance of a device increases when the 
ordered moments of the ferromagnets have antiparallel orientations
on opposite sides of the insulating layer.  Since the relative
moment orientations can be altered by a small
external magnetic field, MTJ  devices can exhibit strong 
magnetoresistance.
Recent progress\cite{moodera56,miyazaki5,gallagher} in the 
fabrication of MTJ devices with large reproducible magnetoresistances,
has led to interest in their possible utility
as digital information storage devices
for nonvolatile random access memories or as read heads. 
Experimenters\cite{moodera56,miyazaki5,gallagher}  
have succeessfully fabricated samples with magnetoresistances
close to theoretical limits proposed by Julli\`{e}re\cite{julliere,ourprl}.
The Julli\'{e}re model ignores the spin-bottleneck
effects which play an important role in studies of 
giant magnetoresistance in magnetic-metal multilayers\cite{gmr}
and in closely related spin-injection experiments\cite{sie,hershfield}. 
In MTJ devices for which the resistance of the ferromagnetic 
electrodes is much smaller than the resistance of  the
tunneling barrier, the spin-bottleneck effect indeed plays 
a minor role.  For modern random access memory 
applications, however, it is important to make MTJ devices with much  
smaller overall resistance to enable operation at fast read access 
times\cite{daughton}.  The desirability of low-resistance microstructured
tunnel junctions\cite{gallagher} raises the issue which 
we address here.  

In this paper we propose a simple electrical-circuit model for 
spin-bottleneck contributions to the resistance and magnetoresistance
of MTJ devices.  
We use the model to evaluate the resistance  of the device,  
for both parallel ($R^{P}$) and antiparallel ($R^{A}$) 
ordered moment orientations. 
An important figure of merit for MTJ devices is the magnetoresistance
defined by 
\begin{equation} 
MR \equiv \frac{R^{A} - R^{P}}{R^{A}}. 
\label{mrdefn}
\end{equation} 
We show that, while both $R^{P}$ and $R^{A}$ 
increase due to the spin-bottleneck effect,
the magnetoresistance may increase or decrease.
We have established the accuracy of the circuit model,
by comparing its predictions with those resulting from 
solutions of the coupled spin-up and spin-down transport
equations for the inhomogeneous device.  
We find that the circuit model gives remarkably good 
estimates for the magnetoresistance of the device, even at a quantitative level.

\begin{figure}[b]
\epsfxsize=3in
\centerline{\epsffile{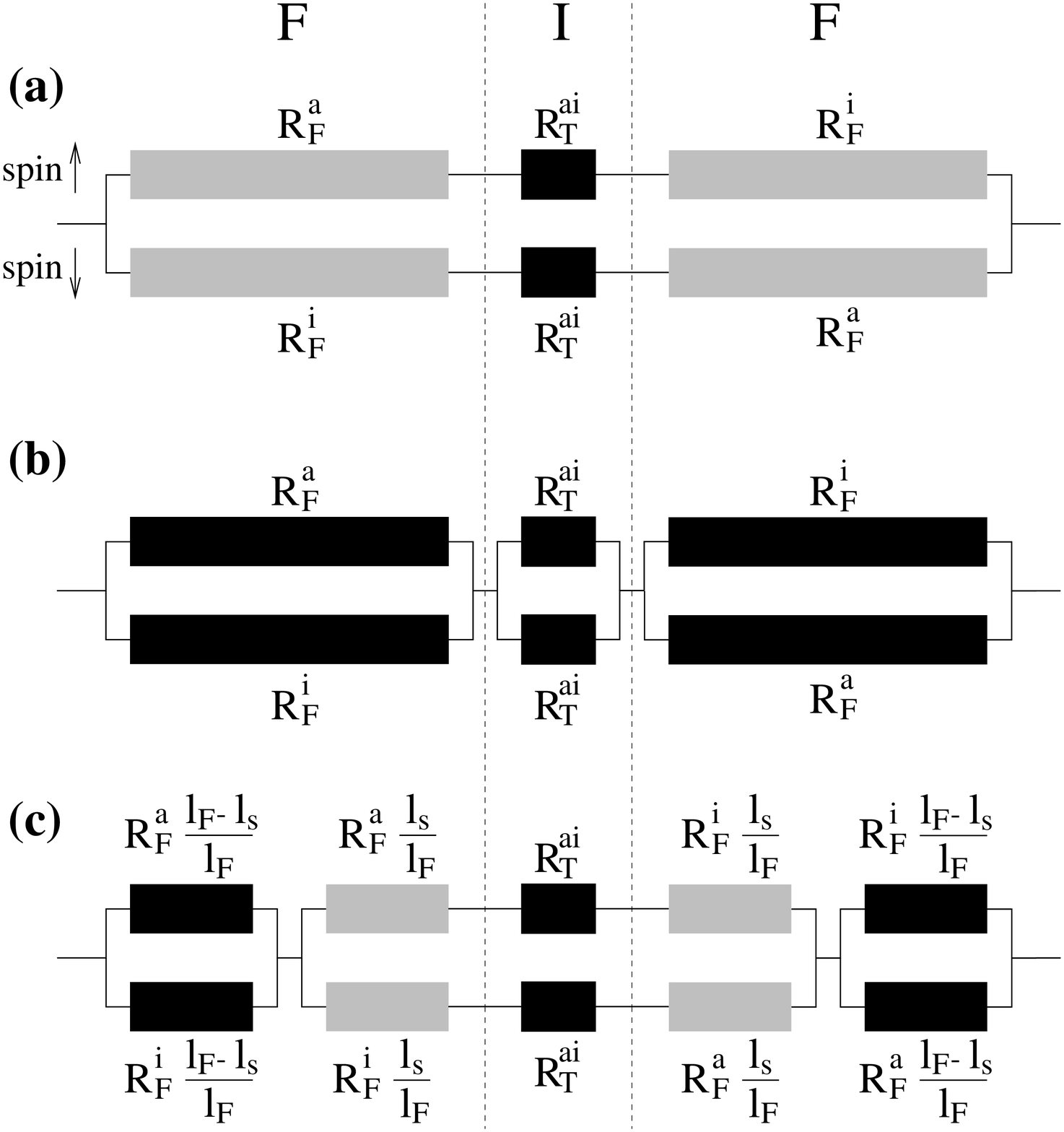}}

\vspace*{.3cm}

\caption{Circuit model for a ferromagnet-insulator\--ferromagnet 
(F-I-F) MTJ device with antiparallel ordered moment 
orientations of the ferromagnets.  Each spin channel has 
resistance contributions from the two electrodes and from the 
tunnel junction.  In our convention the up-spin
channel corresponds to minority spins and the down-spin
channel to majority-spins on the right-hand
side of the junction.  In this configuration the 
tunnel junction resistances in the two spin channels are equal.
In panel (a) there is no spin relaxation and up-spin and 
down-spin channels carry current in parallel.  In panel
(b), spin relaxation is fast enough to maintain equal 
electrochemical potentials for up and down spins
throughout the device.  In panel (c) equilibrium between
up and down spins is established when the distance from
the tunnel junction exceeds the spin-diffusion length of 
the ferromagnetic electrode, $l_s$.  Light shading is 
used for electrode resistors in which up and down spins
are driven from local equilibrium by the inhomogeneous 
spin-dependent transport coefficients of the device.
}
\label{acircuit}
\end{figure}

\begin{figure}[b]
\epsfxsize=3in
\centerline{\epsffile{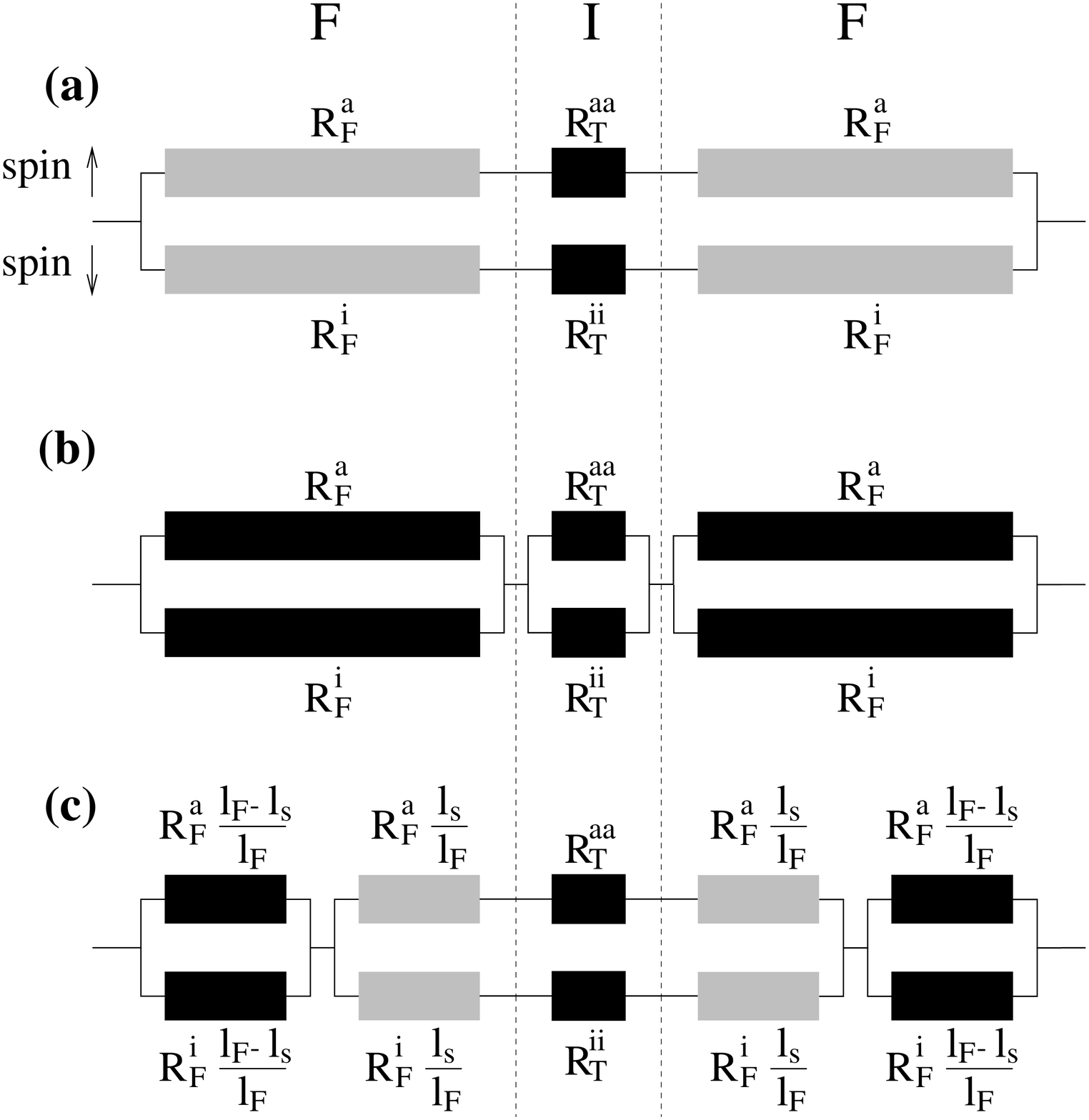}}

\vspace*{.3cm}

\caption{Circuit model for a ferromagnet-insulator\--ferromagnet 
(F-I-F) MTJ device with parallel ordered moment 
orientations.  In this configuration the two spin channels have 
distinct tunnel junction resistances.  The up-spin
channel corresponds to majority spins and the down-spin
channel to minority-spins on the both sides of the junction.  
Light shading is used to signal a spin bottleneck as in 
Fig.~\ref{acircuit}. 
}
\label{pcircuit}
\end{figure}

Our circuit model starts from the recognition\cite{wolfarthreview}
that the two spin-orientations in the ferromagnetic electrodes 
provide two channels which carry current through the 
system in parallel.  If the number of up and down spin
electrons were separately conserved throughout the device,
it would be described by a circuit model with 
up and down spin channels in parallel and three resistors
in series for each channel as illustrated in the top 
panels of Fig.~\ref{acircuit} and \ref{pcircuit}.  To simplify notation 
we assume here that the ferromagnetic electrodes are 
identical in length ($l_F$) and crossection ($A$) and made from
the same material.  When ordered moments are parallel,
the up-spin channel will have the majority spin
electrode resistance, and the down-spin channel will have the minority spin
electrode resistance, on both sides of the tunnel junction.  For 
antiparallel ordered moment orientations, we adopt a convention
where the up-spins are in the majority on the left side of the junction
and in the minority on right side.  In the Julli\`{e}re
model\cite{julliere,ourprl}, 
which appears to be accurate for most materials of interest, 
the tunnel-junction conductance for each spin channel is proportional 
to the product of factors, one for each electrode,
which are dependent on both the density-of-states and the character
of the wavefunctions of that spin at the Fermi energy.
Writing these factors symbolically as $t_a N_a$ and $t_i N_i$ 
for m{\it a}jority and m{\it i}nority spins, respectively, 
we see that when the ordered moment orientations are antiparallel
the tunneling resistances in the two-spin channels are equal:
$R_T^{A,\uparrow}= R_T^{A,\downarrow}
\equiv R_T^{ai}=(t_aN_at_iN_i)^{-1}$. 
For parallel magnetizations $R_T^{P,\uparrow}\equiv R_T^{aa}=(t_aN_a)^{-2}$
and $R_T^{P,\downarrow}\equiv R_T^{ii}=(t_iN_i)^{-2}$.
(In these equations $N_{a}$ and $N_{i}$ are the 
majority and minority spin densities of states and the factors 
$t_{a}$ and $t_{i}$ account for the dependence of tunneling 
amplitudes on wavefunction character.)  
Julli\`{e}re's formula for the magnetoresistance, which in our
notation takes the form 
\begin{equation}
MR=\frac{(t_aN_a - t_iN_i)^2}{t_a^2N_a^2 + t_i^2N_i^2},
\end{equation}
results from the circuit model when the electrode resistances are 
negligible.

For finite electrode resistances and independent conduction
in up and down spin channels, we see from the circuit model
that the local electrochemical potentials of up and down 
spins differ, most strongly at the tunnel junction.
Whenever this occurs, spin-flip processes in the ferromagnet which permit the 
spin distributions to relax to local equilibrium become important.
Separate charge conservation for up and down spins constrains 
the partitioning of current between channels in different parts
of the device and increases its overall resistance
compared to the case where spins are free to change their spin-state
\cite{gmr,sie,hershfield}.
We refer to the resulting increase in resistance as the 
spin-bottleneck resistance of the device.
In the fast-spin relaxation limit, the spin-bottleneck resistance
vanishes and local equilibrium between 
spins is maintained everywhere in the device.
In the circuit model this can be achieved by shorting the 
two spin-channels on both sides of the junction as illustrated
in panel (b) of Figs.~\ref{acircuit} and~\ref{pcircuit}.  
The use of 
a circuit model to represent the transport properties of the device
is  essentially exact in both zero and fast relaxation limits,
being dependent only on the usual locality assumptions.  
We propose its use in the intermediate case where, as we 
discuss briefly below, local equilibrium between the spin
channels is established only when the distance from the 
junction exceeds the  spin-diffusion length $l_s$.  
This can be represented in the circuit model by shorting 
the two spin-channels in each electrode resistor 
at the point $l_s$ from the tunnel junction as illustrated in 
panel (c) of Figs.~\ref{acircuit} and \ref{pcircuit}.  

The circuit model allows the interplay between spin-dependent tunneling
and spin-bottleneck effects to be addressed in a simple and 
intuitive manner.  Formulas for the overall device resistance 
for both parallel and antiparallel orientations follow immediately
from Figs.~\ref{acircuit} and \ref{pcircuit} using elementary 
circuit analysis rules. 
For the intermediate case, $0< l_s< l_F$, we obtain: 
\begin{eqnarray}
R_J^{A}&=&2\frac{R_F^aR_F^i}{R_F^a+R_F^i}\frac{l_F-l_s}{l_F}+
\frac{1}{2}\left[\frac{l_s}{l_F}\left(R_F^a+R_F^i\right)+R_T^{ai}\right]
\nonumber \\
& & \nonumber \\
R_J^{P}&=&2\frac{R_F^aR_F^i}{R_F^a+R_F^i}\frac{l_F-l_s}{l_F}+
\nonumber \\
& & 
\frac{\left(2R_F^al_s/l_F+R_T^{aa}\right)
\left(2R_F^il_s/l_F+R_T^{ii}\right)}
{2\left(R_F^a+R_F^i\right)l_s/l_F+R_T^{aa}+R_T^{ii}}\; .
\end{eqnarray}
Note that in a MTJ device the spin-bottleneck appears 
even when the ordered moment orientations are parallel.

\vspace{-1cm}

\begin{figure}[b]
\epsfxsize=3.6in
\centerline{\epsffile{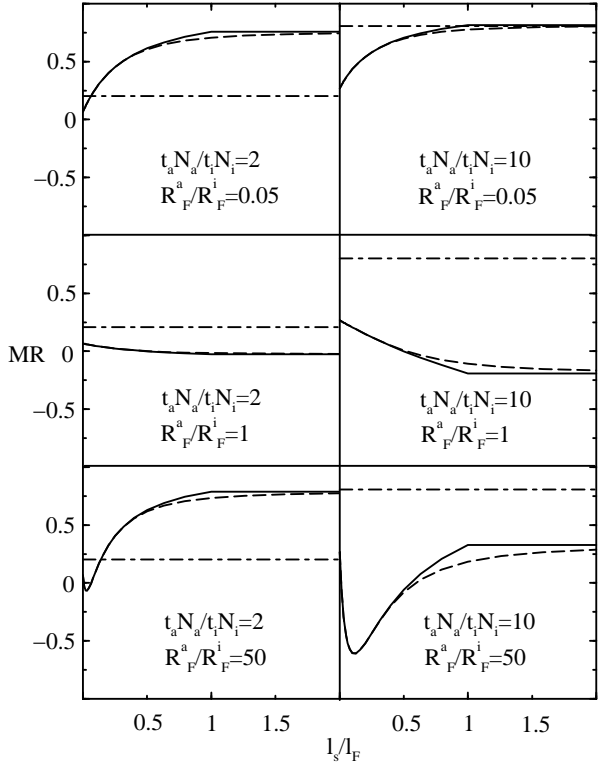}}

\vspace*{-1cm}

\caption{Magnetoresistance as a function of the spin diffusion 
length calculated using the simple circuit model (full lines) and
from solutions of linear transport equations for the inhomogeneous
device (dashed lines).
These results are obtained for a device with comparable
electrode and tunnel junction contributions to the resistance.
The magnetoresistance calculated for zero electrode resistance 
using the Julli\'{e}re formula is indicated by dot-dashed lines.}
\label{mr}
\end{figure}

Typical results for the dependence of overall device magnetoresistance  
on spin diffusion length are shown in Fig.~\ref{mr}.
These results are for the case of a device with comparable 
electrode and tunnel junction resistances; 
$R_F^aR_F^i/(R_F^a+R_F^i)=R_T^{ai}/2$.
Spin-bottleneck contributions to the magnetoresistance 
decline in quantitative importance
along with the electrode contribution to the 
overall resistance, but the trends as a function of 
spin diffusion length are similar to those shown here.  
The left panels in Fig.~\ref{mr} are for the case 
$t_a N_a = 2 t_i N_i $, while the right panels are for
the case $t_a N_a = 10 t_i N_i$.  Neglecting electrode 
resistance, these material parameters correspond to 
$MR=1/5$ and $MR=81/101$ respectively.  The curves 
in Fig.~\ref{mr} are readily
understood using the simple circuit model.   The dependence of overall 
resistance on moment orientation is due completely to the 
central portion of the circuit in which the two spin states
are not in local equilibrium.  For antiparallel orientations this 
portion consists of two identical resistors in parallel with 
resistance $R^{*} = (l_s/l_F)(R_F^a+R_F^i) +R_T^{ai}$.  For  
parallel orientations the two resistors in the central regions are 
not identical and have values $R^{a*} = 2 (l_s/l_F) R_F^a + R_T^{aa}$ and 
$R^{i*} = 2 (l_s/l_F) R_F^i + R_T^{ii}$.  Note that the electrode
contribution to $R^{*}$ is the arithmetic mean of the electrode
contributions to $R^{a*}$ and $R^{i*}$ whereas the tunnel junction
resistance contribution is the geometric mean of the the tunnel
junction contributions to $R^{a*}$ and $R^{i*}$.  The property 
$(R^{*})^2 = R^{a*}R^{i*}$, assuming the Julli\`{e}re formula, 
holds when electrode resistances are neglected and guarantees  
a positive magnetoresistance.  This property is lost in the general
case and $MR$ can be negative.  $MR$ will tend to be 
large and positive when either $R^{a*}$ or $R^{i*}$ is substantially
smaller than $R^{*}$. 

We have considered only cases
where the tunneling conductance is larger for majority spins,
since this appears\cite{meservey,moodera56} 
to be the case for all systems studied
experimentally.  The three panels from top to bottom in 
Fig.~\ref{mr} are for majority spin resistivities, smaller,
equal to and larger than minority spin resistivities.  All
three cases can occur\cite{wolfarthreview} in metallic ferromagnets,
depending on the material and the disorder potential.  For $l_s=0$, 
$R^{A}-R^{P}$ equals its Julli\`{e}re value, and 
$MR$ is reduced only because of the overall resistance increase  
coming from the electrodes.  In both top panels, $R^{a*}/R^{*}$, 
decreases further below $1$ as $l_s$ increases, leading to
an increase in $MR$.  For the case $R_F^{a}= R_F^{i}$ illustrated
in the middle panels, $R^{a*}/R^{*}$ increases toward $1$ as the electrode
contributions are added and $MR$ eventually changes sign
as $l_s$ approaches $l_F$.  As mentioned above $MR$ need not be positive in the 
general case.  In the bottom panels, $R^{a*}/R^{*}$ increases
as $l_s$ increases from $0$, causing $MR$ to change sign,
but for larger $l_s$ the electrode resistance effect  
begins to dominate, $R^{i*}$ becomes much smaller than
$R^{*}$ and $MR$ again becomes positive.  The minimum $MR$ occurs
at smaller $l_s$ in the left panel case, because 
of the smaller difference in tunneling resistances.  Similar 
simple considerations can be used to obtain an intuitive understanding
of the magnetoresistance magnitude and sign for other 
material parameter values.  

The electric circuit model, presented above, allows $MR$ values to
be understood simply, but is based on an approximate representation of 
the space and spin dependent local electrochemical potentials
in the inhomogeneous MTJ device.  To assess its quantitative 
accuracy we have compared its predictions for the overall device 
resistance and magnetoresistance with the results of a linear 
response calculation which assumes only a local conductivity for 
each spin channel\cite{gmr,sie}.
In this calculation, spin-up and spin-down 
electrons carry current in parallel but spin-relaxation
can occur everywhere in the ferromagnetic 
electrodes.  Applying Ohm's low locally, the current densities for each
spin channel are given by 
\begin{equation}
\label{ohm}
j_{\uparrow}=\frac{\sigma_{\uparrow}}{|e|}\frac{d\overline{\mu}_{
\uparrow}}{dx}\; , \;\;
j_{\downarrow}=\frac{\sigma_{\downarrow}}{|e|}\frac{d\overline{\mu}_{
\downarrow}}{dx} \; ,
\end{equation}
where $\overline{\mu}_{\uparrow(\downarrow)}$ is the space and spin
dependent electrochemical\cite{eleccaveat} potential 
and $\sigma_{\uparrow (\downarrow)}=l_F/(AR_F^{\uparrow (\downarrow)})$.
Applying the continuity 
equation to both spin channels and employing a 
relaxation time approximation implies that 
\begin{equation}
\label{cont}
\frac{dj_{\uparrow}}{dx}=-\frac{dj_{\downarrow}}{dx}=
\frac{|e|}{\mu_B}\frac{M(x)-M_{eq}(x)}{\tau_s} \; .
\label{ce}
\end{equation}
The non equilibrium magnetization, $M(x)$, is proportional to the difference
between number of spin up and down electrons, i.e.,
$M(x)= M_0 + \mu_B[N_{\uparrow}\mu_{\uparrow}(x)-
N_{\downarrow}\mu_{\downarrow}(x)]$, where 
$\mu_{\uparrow(\downarrow)}$ is the space and spin dependent
chemical potential, the Fermi energy has been chosen 
as the zero of energy, and $M_0$ is the saturation moment of
the ferromagnet.  $M_{eq}(x)$ is the local quasi-equilibrium
magnetization towards which $M(x)$ relaxes.
If we assume that the exchange-splitting of the 
equilibrium ferromagnetic bands\cite{tobepub} is not
altered by a change in local electron density, it 
follows that $M_{eq}(x) = M_0 + \mu_B(N_{\uparrow} - N_{\downarrow})
[N_{\uparrow}/(N_{\uparrow} + N_{\downarrow}) \mu_{\uparrow}
 + N_{\downarrow}/(N_{\uparrow}+N_{\downarrow}) \mu_{\downarrow}]$ 
and we obtain\cite{hershfield}  
\begin{eqnarray}
M-M_{eq}&=& \frac{2\mu_BN_{\uparrow}N_{\downarrow}}{N_{\uparrow}+N_{\downarrow}}
(\mu_{\uparrow}-\mu_{\downarrow})\nonumber \\
&=& \frac{2\mu_BN_{\uparrow}N_{\downarrow}}{N_{\uparrow}+N_{\downarrow}}
(\overline{\mu}_{\uparrow}-\overline{\mu}_{\downarrow})\, .
\label{m-meq}
\end{eqnarray}

Eqs.~(\ref{ohm}) and (\ref{ce}) with the
relaxation model~(\ref{m-meq}) implies  
electrochemical potential differences between 
up and down spins which decay exponentially 
over a length scale $l_s$ inside the electrodes
\begin{equation}
l_s=\sqrt{\frac{\tau_s}{2e^2}\frac{N_a+N_i}{N_aN_i}
\frac{\sigma_{\uparrow}\sigma_{\downarrow}}{\sigma_{\uparrow}+
\sigma_{\downarrow}}} \; .
\end{equation}
These equations can be solved to determine the partitioning
of current between spin channels throughout the device. 
The tunnel barrier enters the 
calculations through the assumption of  
electrochemical potential drops of $\delta\overline{
\mu}_{\uparrow}=AR_T^{\uparrow}j_{\uparrow}$ and $\delta\overline{ 
\mu}_{\downarrow}=AR_T^{\downarrow}j_{\downarrow}$.
The other boundary conditions, needed to
solve Eqs.(~\ref{ohm}) and (\ref{cont}), 
express the continuity of the current of each spin in
the tunneling barrier and the recovery of equilibrium spin polarization
at the ends of the ferromagnetic electrodes\cite{cond}.  The circuit model
captures the essence of these equations in a simple way.
A comparison between the circuit model and the macroscopic transport 
theory outlined above is shown in Fig.~\ref{mr}. 
The two models agree almost
exactly for $l_s$ up to one half of $l_F$. The small deviation at larger
$l_s$ results from the fact that circuit model has a cusp at $l_s=l_F$
and so is uncapable of describing the smooth saturation of the
MR curve.  The good overall quantitative agreement 
between the two models justifies the use of 
the equivalent circuits of Figs.~\ref{acircuit} and \ref{pcircuit}
to understand the magnetoresistance of MTJ 
devices with low overall resistances.  

The authors acknowledge helpful interactions with
W.H. Butler, J.F. Cooke, J.S. Moodera, I. Schuller, 
and R.S. Sooryakumar.  This work was supported by the National Science
Foundation under grants DMR-9714055 and INT-9602140, 
by the Ministry of Education of the Czech Republic under grant ME-104,
and by the Grant Agency of the Czech Republic under grant 202/98/0085.


\begin{references}
\bibitem{eleccaveat} The electrochemical potential and 
chemical potential are distinguished in our terminology by
the inclusion in the former of the electric potential 
which is presumed to be slowly varying and is responsible for  
rigid bending of the ferromagnet's bands.
\bibitem{cond} This boundary condition is based on the assumption
that the spin-diffusion length in the contacts made to the MTJ device are
much smaller than $l_F$. If this is not the case the circuit model
can be generalized, in a straightforward way, to account spin-relaxation
in the contacts.
\end{references}
\end{document}